\documentclass[superscriptaddress,onecolumn,secnumarabic,
amssymb,amsmath,nobibnotes,aps,prd,showkeys,showpacs,nofootinbib]{revtex4}

\usepackage[latin1]{inputenc}
\usepackage{graphicx}
\usepackage[english]{babel}

\usepackage{amsmath}
\usepackage{amssymb}
\usepackage{amsfonts}
\usepackage{colordvi}
\usepackage{psfrag}
\usepackage{color}

\begin{document}

\def\cL{{\cal L}}
\def\be{\begin{equation}}
\def\ee{\end{equation}}
\def\bea{\begin{eqnarray}}
\def\eea{\end{eqnarray}}
\def\beq{\begin{eqnarray}}
\def\eeq{\end{eqnarray}}
\def\tr{{\rm tr}\, }
\def\nn{\nonumber \\}
\def\e{{\rm e}}

%\begin{document}

\def\bef{\begin{figure}}
\def\eef{\end{figure}}
\newcommand{\ans}{ansatz }
\newcommand{\eeqn}{\end{eqnarray}}
\newcommand{\bd}{\begin{displaymath}}
\newcommand{\ed}{\end{displaymath}}
\newcommand{\mat}[4]{\left(\begin{array}{cc}{#1}&{#2}\\{#3}&{#4}
\end{array}\right)}
\newcommand{\matr}[9]{\left(\begin{array}{ccc}{#1}&{#2}&{#3}\\
{#4}&{#5}&{#6}\\{#7}&{#8}&{#9}\end{array}\right)}
\newcommand{\matrr}[6]{\left(\begin{array}{cc}{#1}&{#2}\\
{#3}&{#4}\\{#5}&{#6}\end{array}\right)}
\newcommand{\cvb}[3]{#1^{#2}_{#3}}
\def\lsim{\raise0.3ex\hbox{$\;<$\kern-0.75em\raise-1.1ex
e\hbox{$\sim\;$}}}
\def\gsim{\raise0.3ex\hbox{$\;>$\kern-0.75em\raise-1.1ex
\hbox{$\sim\;$}}}
\def\abs#1{\left| #1\right|}
\def\simlt{\mathrel{\lower2.5pt\vbox{\lineskip=0pt\baselineskip=0pt
           \hbox{$<$}\hbox{$\sim$}}}}
\def\simgt{\mathrel{\lower2.5pt\vbox{\lineskip=0pt\baselineskip=0pt
           \hbox{$>$}\hbox{$\sim$}}}}
\def\unity{{\hbox{1\kern-.8mm l}}}
\newcommand{\eps}{\varepsilon}
\def\ep{\epsilon}
\def\ga{\gamma}
\def\Ga{\Gamma}
\def\om{\omega}
\def\omp{{\omega^\prime}}
\def\Om{\Omega}
\def\la{\lambda}
\def\La{\Lambda}
\def\al{\alpha}
\newcommand{\ov}{\overline}
\renewcommand{\to}{\rightarrow}
\renewcommand{\vec}[1]{\mathbf{#1}}
\newcommand{\vect}[1]{\mbox{\boldmath$#1$}}
\def\tm{{\widetilde{m}}}
\def\mcirc{{\stackrel{o}{m}}}
\newcommand{\Dm}{\Delta m}
\newcommand{\dm}{\varepsilon}
\newcommand{\tanb}{\tan\beta}
\newcommand{\nbar}{\tilde{n}}
\newcommand\PM[1]{\begin{pmatrix}#1\end{pmatrix}}
\newcommand{\up}{\uparrow}
\newcommand{\down}{\downarrow}
\def\omE{\omega_{\rm Ter}}
%
%%%%%%%%%%     mauri    %%%%%%%%%%%%%%%%%%%%%%%%%%%%%%%%%

\newcommand{\Dsusy}{{susy \hspace{-9.4pt} \slash}\;}
\newcommand{\DCP}{{CP \hspace{-7.4pt} \slash}\;}
\newcommand{\mc}{\mathcal}
\newcommand{\gr}{\mathbf}
\renewcommand{\to}{\rightarrow}
\newcommand{\gtc}{\mathfrak}
\newcommand{\wh}{\widehat}
\newcommand{\br}{\langle}
\newcommand{\kt}{\rangle}

%%%%%%%%%%%%%%%%%%%%%%%%%%%%%%%%%%%%%%%%%%%%%%%%%%%%%%%%%%

% barbara Ricci  %definizione di minore e maggiore simile
\def\lsim{\mathrel{\mathop  {\hbox{\lower0.5ex\hbox{$\sim$}
\kern-0.8em\lower-0.7ex\hbox{$<$}}}}}
\def\gsim{\mathrel{\mathop  {\hbox{\lower0.5ex\hbox{$\sim$}
\kern-0.8em\lower-0.7ex\hbox{$>$}}}}}
%%%%%%%%%%%%%%%%%%%%%%%%%%%%%%%%%%

\def\nn{\\  \nonumber}
\def\de{\partial}
\def\brf{{\mathbf f}}
\def\bbf{\bar{\bf f}}
\def\bF{{\bf F}}
\def\bbF{\bar{\bf F}}
\def\bA{{\mathbf A}}
\def\bB{{\mathbf B}}
\def\bG{{\mathbf G}}
\def\bI{{\mathbf I}}
\def\bM{{\mathbf M}}
\def\bY{{\mathbf Y}}
\def\bX{{\mathbf X}}
\def\bS{{\mathbf S}}
\def\bb{{\mathbf b}}
\def\bh{{\mathbf h}}
\def\bg{{\mathbf g}}
\def\bla{{\mathbf \la}}
\def\bmu{\mathbf m }
\def\by{{\mathbf y}}
\def\bmu{\mbox{\boldmath $\mu$} }
\def\bsig{\mbox{\boldmath $\sigma$} }
\def\bunity{{\mathbf 1}}
\def\cA{{\cal A}}
\def\cB{{\cal B}}
\def\cC{{\cal C}}
\def\cD{{\cal D}}
\def\cF{{\cal F}}
\def\cG{{\cal G}}
\def\cH{{\cal H}}
\def\cI{{\cal I}}
\def\cL{{\cal L}}
\def\cN{{\cal N}}
\def\cM{{\cal M}}
\def\cO{{\cal O}}
\def\cR{{\cal R}}
\def\cS{{\cal S}}
\def\cT{{\cal T}}
\def\eV{{\rm eV}}

\title{The fate of Schwarzschild-de Sitter Black Holes in $f(R)$ gravity}

\author{Andrea Addazi}

\affiliation{ Dipartimento di Fisica,
 Universit\`a di L'Aquila, 67010 Coppito AQ, Italy}
 
 \affiliation{Laboratori Nazionali del Gran Sasso (INFN), 67010 Assergi AQ, Italy}
 
\author{Salvatore  Capozziello}
\email[Corresponding author: ]{capozziello@na.infn.it}
\affiliation{Dipartimento di  Fisica, Universit\`{a} di Napoli "Federico II",
 Compl. Univ. di Monte S. Angelo, Edificio G, Via Cinthia, I-80126, Napoli, Italy,}
\affiliation{Istituto Nazionale di  Fisica Nucleare (INFN) Sez. di Napoli, Compl. Univ. di Monte S. Angelo,
Edificio G, Via Cinthia, I-80126, Napoli, Italy,} \affiliation{Gran
Sasso Science Institute (INFN),  Viale F. Crispi, 7, I-67100,
L'Aquila, Italy.}

\date{\today}

%\vspace{1cm}
\begin{abstract}
The semiclassical effects of 
 antievaporating black holes can be discussed in the framework of $f(R)$ gravity.
In particular,  the Bousso-Hawking-Nojiri-Odinstov  antievaporation instability 
 of degenerate Schwarzschild-de Sitter black holes
 (the so called Nariai space-time)  
 leads to a dynamical increasing of black hole horizon in $f(R)$ gravity.
 This phenomenon  causes the following transition: emitting marginally trapped surfaces 
  become space-like surfaces before the effective 
 Bekenstein-Hawking emission time. As a consequence, 
 Bousso-Hawking thermal radiation cannot be emitted 
 in an antievaporating Nariai black hole. Possible implications in cosmology and black hole physics are also discussed. 

%%%

\end{abstract}
\pacs{04.50.Kd,04.70.-s, 04.70.Dy, 04.62.+v, 05.,05.45.Mt}
\keywords{Alternative theories of gravity,  black hole physics, quantum black holes, quantum field theory in curved space-time}

\maketitle

\section{Introduction}

In \cite{Nojiri:2013su}, it is discussed the antievaporation
effect for degenerate Schwarzschild-de Sitter black holes in the context of $f(R)$ gravity. 
 The authors  demonstrated 
 that the black hole radius  increases because of extra 
 gravitational contribution from the extended 
 gravitational action. In \cite{Nojiri:2014jqa}, the generalization to 
  charged black holes is studied, showing the same effect. 
 An analogous phenomena was previously studied by Bousso and Hawking \cite{Bousso:1997wi}. Antievaporation is also discussed in the contest of bigravity in \cite{Katsuragawa:2014hda}.
 
 In this paper, we take into account  implications of such a phenomena 
 in a quantum semiclassical regime. 
One could expect that also in this case Bekenstein-Hawking 
 radiation would be emitted by black holes in $f(R)$-gravity; {\it i.e}
 Bousso-Hawking-Nojiri-Odintsov  antievaporation and Bekenstein-Hawking evaporation 
  would be both present in $f(R)$ black holes. 
 However, we shall demonstrate that Bekenstein-Hawking
 evaporation is completely suppressed by 
Bousso-Hawking-Nojiri-Odintsov antievaporation in the context of Nariai space-time (Nariai black holes), for a large class of $f(R)$ gravity models. 
This result may appear surprising.  However, let us remark that the derivation of Bekenstein-Hawking
 radiation is  based on a non-dynamical background metric. 
 In other words, they consider black holes
 in the limit of infinite mass, heat capacity 
 and static event horizon. 
 As a consequence, their argument cannot be 
 rigorously applied in a dynamical case. 
 We are going to  show that the Bousso-Hawking-Nojiri-Odintsov antievaporation 
 introduces an extra focusing term into the  
Raychaduri equation. This implies that an emitting marginally trapped surface 
  transits from a time-like to a space-like surface in a short range of time. 
  In other words, a time-like emitting horizon surface is almost suddenly   
  trapped into the black hole space-like interior. 
  However, the Bekenstein-Hawking radiation cannot be emitted from space-like surface. This  result is well known in literature:
  the original Bekenstein and Hawking argument, using Bogolubov coefficients
  \cite{Bekenstein,Hawking},
  tunnel effect calculations \cite{tunneling}, or eikonal approximations \cite{eikonal}
  lead to this conclusion.  
As a consequence, the Bekenstein-Hawking radiation is suppressed in an antievaporating space-time.
As a shown in
\cite{Ellis:2013oka,Firouzjaee:2014zfa,Firouzjaee:2015bqa}, a  pair creation in a dynamical space-time
 violates the energy conservation, {\it i.e} the (quantum) stress-energy  tensor conservation. 
We will show below how similar argument can be applied for  $f(R)$ gravity black holes. 

The  paper is organized as follows:
in Section 2, we briefly review the main results on  Bouss-Hawking-Nojiri-Odintsov antievaporation;
 Section 3 is devoted to the  mechanism    of Bekenstein-Hawking radiation suppression
starting from the path integral formalism, developed  in Section 3.1. 
In Section 3.2, our main argument is discussed. 
In Section 4, we  draw conclusions and give outlooks on possible 
cosmological implications.

\section{Bousso-Hawking-Nojiri-Odintsov antievaporation instability}  

Let us review now some basic aspects 
of antievaporation in $f(R)$ gravity.
Details of this derivation can be found in \cite{Nojiri:2013su}. 
Let us consider the $f(R)$ gravity action 
\be \label{action}
{I}=\frac{1}{16\pi}\int d^{4}x\sqrt{-g}f(R)+S_{m}
\ee
written in units $G_{N}=c=1$.
The field equations  are 
\begin{equation}
f'(R)R_{\mu\nu}-\frac{1}{2}f(R) g_{\mu\nu}-\left[\nabla_{\mu} \nabla_{\nu} - g_{\mu\nu} \Box\right]f'(R)= 8\pi T_{\mu \nu },
\label{EoM}
\end{equation}
where $T_{\mu\nu}$ is stress-energy tensor of matter.
%\be \label{EoM}
%\frac{1}{2}g_{\mu\nu}f(R)-f'(R)R_{\mu\nu}+\nabla_{\mu}\nabla_{\nu}f'(R)=-8\pi T^{m}_{\mu\nu}
%\ee
For $T^{m}_{\mu\nu}=0$, assuming  the Ricci tensor  covariantly constant and proportional to $g_{\mu\nu}$, the field equations 
 reduced to 
\be \label{EoM2}
f(R)-\frac{1}{2}Rf'(R)=0\,.
\ee
The Nariai space-time is a solution of Eq.\eqref{EoM2},
having the following form:
\be \label{Nariai}
ds^{2}=\frac{1}{\mathcal{M}^{2}}\left[\frac{1}{\cos h^{2}x}(dx^{2}-dt^{2})+d\Omega^{2}\right]\,,
\ee
where $\mathcal{M}$ is a  mass scale, and the $d\Omega^{2}$ is the 
solid angle on a 2-sphere $d\Omega^{2}=d\theta^{2}+\sin^{2}\theta d\phi^{2}$. 
The Ricci scalar of Nariai space-time is 
 a constant being  $R=4\mathcal{M}^{2}$.
Dynamical aspects are achieved by
 perturbing  the Nariai space-time. Let us  assume  the  general expression for the metric
\be \label{pertNariai}
ds^{2}=e^{2\rho(x,t)}(dx^{2}-dt^{2})+e^{-2\phi(x,t)}d\Omega^{2}
\ee
where
\be
\rho=-\ln(\mathcal{M}\cosh x)+\delta \rho)\,.\qquad
\phi=\ln \mathcal{M}+\delta \phi\,.
\ee
Substituting these expressions into the field equations, 
a  set of equations in $\delta \rho,\delta \phi$ is obtained.
The horizon radius has the form 
\be \label{Horizon}
r_{H}=\frac{1}{\mathcal{M}}e^{-\phi_{0}\cosh^{2}\beta t}\,,
\ee
with the parameterization
\be
\delta \phi=\phi_{0}\,\cosh \omega t \,\cosh^{\beta} x=\phi_{0}\cosh^{2}\beta t\,,
\ee
and $\omega,\beta$ real parameters related each other by the field equations and the 
horizon definition $g^{\mu\nu}\nabla_{\mu}\phi \nabla_{\nu}\phi=0$. 
If $\phi_{0}<0$, $r_{H}$ increases,
{\it i.e}  antievaporation phenomena happen. 
\footnote{Another branch of solution, corresponding to $\omega,\beta$ complex parameters,
leads to more exotic solutions: an infinite number of evaporating and antievaporating horizons is conjectured. 
This case  will not be discussed in this paper.  }

According to \cite{Nojiri:2013su}, let us consider  a class of $f(R)$ models 
\be \label{fRc}
f(R)=\frac{R}{2\kappa^{2}}+f_{2}R^{2}+f_{0}\mathcal{M}^{4-2n}R^{n}\,.
\ee
In this case, 
antievaporation  occurs 
for $n>2$ and $\zeta>-n/2\,\,\, \& \,\,\,\zeta<\frac{32-19n}{6}$
and for $n<2$ and $\zeta<-\frac{n}{2}\,\,\,\& \,\,\,\zeta>\frac{32-19 n}{6}$;
where 
\be
\zeta=\frac{f_{2}}{n-1}[2(n-2)n^{2}]^{\frac{2-n}{1-n}}f_{0}^{\frac{1}{1-n}}\mathcal{M}^{\frac{4-n}{1-n}}\,.
\ee
As a consequence, antievaporation is happening 
in a large space of  parameters.
In other gravity theories, extra polynomial terms in the Newtonian potential 
can lead to the destabilizations of geodetics, as shown in \cite{Addazi:2014mga}.

\section{The suppression of Bekenstein-Hawking radiation}

Let us  discuss  now the suppression of
Bekenstein-Hawking radiation in $f(R)$-gravity-Nariai black holes considering theoretical approaches where such a suppression comes out\footnote{Our derivation is inspired to the  papers \cite{Ellis:2013oka,Firouzjaee:2014zfa,Firouzjaee:2015bqa}.}.

\subsection{The Path integral approach}

In general,
the path integral over all Euclidean metrics and matter fields $\phi_{i},\psi_{j},A^{\mu}_{k},..$
is 
\be \label{formaly}
Z_{E}=\int \mathcal{D}g\mathcal{D}\phi_{i}\mathcal{D}\psi_{j}\mathcal{D}A^{\mu}_{k}e^{-I[g,\phi_{i},\psi_{j},A^{\mu}_{k},...]}
\ee
where $g$ is the Euclidean metric tensor. 
Adopting a  semiclassical limit of  General Relativity, the
leading terms in the action are
\cite{Euclidean} 
\be \label{I}
I_{E}=-\int_{\Sigma}\sqrt{g}d^{4}x\left(\mathcal{L}_{m}+\frac{1}{16\pi}R\right)+\frac{1}{8\pi}\int_{\partial \Sigma}\sqrt{h}d^{3}x(K-K^{0})
\ee
where $\mathcal{L}_{m}$ is the matter Lagrangian  
\be \mathcal{L}_{m}=\frac{Y^{ii'}}{2}g_{\mu\nu}\partial \phi^{i\mu}\partial \phi^{i'\nu}+...\ee
 $K$ is the trace of the curvature induced on the boundary $\partial \Sigma$
of the region $\Sigma$ considered, $h$ is the metric induced on the boundary $\partial \Sigma$, and 
$K^{0}$ is the trace of the induced curvature embedded in flat space.
The last term is a contribution from the boundary.
We consider infinitesimal perturbations of matter and metric
as $\phi=\phi_{0}+\delta \phi$,
$A=A^{0}+ \delta A$, (...)
and $g=g_{0}+\delta g$, 
so that 
\be I[\phi,A,...,g]=I[\phi_{0},A_{0},..g_{0}]+I_{2}[\delta \phi,\delta A,...\delta g]+\mbox{higher orders}\,,\ee
\be I_{2}[\delta \phi,\delta A,..,\delta g]=I_{2}[\delta \phi,\delta A,...]+I_{2}[\delta g]\,.\ee
\be \label{Z}
log Z=-I[\phi_{0},A^{0},...,g_{0}]+log \int \mathcal{D}\delta \phi \mathcal{D}\delta A(...) \mathcal{D}\delta g e^{-I_{2}[\delta g,\delta \phi,\delta A,...]}
\ee

In an Euclidean Schwarzschild solution, 
the metric has a time dimension compactified on a circle $S^{1}$, with periodicity $i\beta$,
and 
$$\beta=T^{-1}=8\pi M$$
 Here $T,M$ are the Bekenstein-Hawking temperature and mass respectively. 
The Euclidean Schwarzschild  metric has the  form 
\be \label{form1}
ds_{E}^{2}=\left(1-\frac{2M}{r}\right)d^{2}\tau+\left(1-\frac{2M}{r}\right)dr^{2}+r^{2}d\Omega^{2}
\ee
A convenient change of coordinates 
$$x=4M\sqrt{1-\frac{2M}{r}}$$
leads to 
\be \label{ES}
ds_{E}^{2}=\left(\frac{x}{4M}\right)^{2}+\left(\frac{r^{2}}{4M^{2}}\right)^{2}dx^{2}+r^{2}d\Omega^{2}
\ee
Eq.\eqref{ES} has no more a singularity in $r=2M$. 
The boundary 
$\partial \Sigma$ is $S^{1}\times S^{2}$ where $S^2$ has a  
a conveniently fixed radius $r_{0}$.
The path integral becomes  
a partition function of a (canonical) ensamble, 
with an Euclidean time related to the temperature $T=\beta^{-1}$.
The leading contribution to the path integral is 
\be \label{dom}
Z_{ES}= e^{-\frac{\beta^{2}}{16\pi}}
\ee
Contributions to this term are only coming 
from surface terms in the gravitational action,
{\it i.e} bulk geometry not contributes to Eq.\ref{dom}. 

The average energy (or internal energy) is 
\be \label{average}
\langle E \rangle=-\frac{d}{d\beta}(\log Z)=\frac{\beta}{8\pi} 
\ee
On the other hand, the free energy $F$ is related to $Z$
as 
\be \label{usuallyd}
F=-T\rm \log Z
\ee
Finally the entropy is 
\be \label{SFU}
S=\beta(F-\langle E \rangle)
\ee
As a consequence, Bekenstein-Hawking radiation can be related to the partition function as follows: 
\be \label{S1}
S=\beta(\log Z-\frac{d}{d\beta}(\log Z))=\frac{\beta^{2}}{16\pi}=\frac{1}{4}A\,.
\ee
We can reformulate the Euclidean approach in $f(R)$ gravity.
%Through a suitable transformation, we can  map 
%$f(R)$ gravity into a scalar-tensor theory adopting  the so called O'Hanlon representation (see \cite{Capozziello:2011et} for details).
The  action, in semiclassical regime,  is now
%\be \label{NRA}
%I=-\frac{1}{16\pi}\int_{\Sigma} d^{4}x\sqrt{g}\left(
%f(\phi)+f'(\phi)(R-\phi)\right)-\frac{1}{8\pi}\int_{\partial \Sigma}d^{3}x\sqrt{h}f'(\phi)(K-K_{0})
%\ee
%that can be remap to the corresponding $f(R)$-gravity action as
\be \label{PluggingBack}
I=-\frac{1}{16\pi}\int d^{4}x\sqrt{-g}f(R)-\frac{1}{8\pi}d^{4}x\sqrt{h}f'(R)(K-K_{0})
\ee
Let us assume a generic spherical symmetric static solution for $f(R)$-gravity 
with an Euclidean periodic time $\tau \rightarrow \tau+\beta$ 
where $\beta=8\pi M$,
\be \label{Euc}
ds_{E}^{2}=J(r)d\tau^{2}+J(r)^{-1}dr^{2}+r^{2}d\Omega^{2}\,.
\ee
As in General Relativity, the leading contribution is zero from the bulk geometry.
On the other hand, the boundary term has a non-zero contribution.
One can evaluate the boundary integral considering 
suitable surface $\partial \Sigma$. 
In this case the obvious choice is a $S^{2}\times S^{1}$ surface with 
with radius $r$ of $S^{2}$. 
We obtain 
\be \label{KKz}
\int_{\partial \Sigma}d^{3}x\sqrt{h}f'(R)(K-K_{0})=f'(R_{0})\int_{\partial \Sigma}d^{3}x\sqrt{h}(K-K_{0})=8\pi\beta r-12\pi\beta M-8\pi\beta r\sqrt{1-\frac{r_{S}}{r}}\,,
\ee
where $r_{S}=2M$ and $R_{0}$ is the scalar curvature of the classical black hole background.
In the limit of $r\rightarrow \infty$, the resulting action, partition function and entropy are
\be \label{ActionFinal}
I=f'(R_{0})\beta^{2},\,\,\,\,\,Z_{E}=e^{-f'(R_{0})\beta^{2}},\,\,\,\,\,S=16\pi f'(R_{0})\frac{A}{4}\,.
\ee 
See also \cite{Dyer:2008hb}.
This result seems in antithesis with our statements in the introduction:
Eqs.(\ref{ActionFinal}) leads to a Bekenstein-Hawking-like radiation. 
In fact, as mentioned, a Nariai solution is nothing else but 
a Schwarzschild-de Sitter one
with $J(r)=1-J(r)_{Schwarzschild}-\frac{\Lambda}{3}r^{2}$, with a black hole 
radius $r\simeq H^{-1}$ (limit of black hole mass $M\rightarrow \frac{1}{3}\Lambda^{-1/2}$),
with mass scale $\mathcal{M}=\Lambda$. 
However, the result (\ref{ActionFinal}) is based on a strong assumption on the metric (\eqref{Euc}):
 the gravitational action does not  to give  dynamical evolution. 
For example, in Nariai solution obtained by Nojiri and Odintsov in 
$f(R)$-gravity, $J(r,t)$ is also a function of time: the mass parameter 
is a function of time $r_{S}(t)$. 
%As a consequence, the analysis performed
%here does not work.
% Let us note that this is also an example 
%of a Birkhoff's theorem violation in $f(R)$ gravity.
%Birkhoff's theorem guarantees that  
%a spherical symmetric Schwarzschild-like solution 
%will be a stable, stationary and static one.
As a consequence, the result got in this section has to be considered 
with caution: Eq.\eqref{ActionFinal} can be applied if and only if one has a spherically symmetric stationary and static solution in  $f(R)$ gravity.  This means that  because the Birkhoff theorem cannot be, in  general,  applied
to  $f(R)$ gravity \cite{Capozziello:2011wg}, non-static cases can be considered.

Let us also comment that the same entropy in (\ref{ActionFinal})
can be obtained by the Wald entropy charge integral. 
The Wald entropy is 
\be \label{WE}
S_{W}=-2\pi \int_{S^{2}} d^{2}x\sqrt{-h^{(2)}}\left( \frac{\delta \mathcal{L}}{\delta R_{\mu\nu\rho\sigma}}\right)_{S^{2}}\hat{\epsilon}_{\mu\nu}\hat{\epsilon}_{\rho\sigma}=\frac{A}{4G_{eff}}
\ee
where $\hat{\epsilon}$ is the antisymmetric binormal vector 
to the surface $S^{2}$
and 
\be \label{Geff}
(2\pi G_{eff})^{-1}=-\left(\frac{\delta \mathcal{L}}{\delta R_{\mu\nu\rho\sigma}}\right)_{S^{2}}\hat{\epsilon}_{\mu\nu}\hat{\epsilon}_{\rho\sigma}
\ee
leading to $G_{eff}=G/f'(R_{0})$ \cite{Briscese:2007cd}.

However, again, this result can be applied if and only if the spherically symmetric solution is static.
As discussed above, this is not the case of Nariai black holes in $f(R)$ gravity. 
This point deserves a further discussion.

The Euclidean path integral approach is supposing an Euclidean black hole 
inside an ideal box in thermal equilibrium with it. Furthermore, the  thermodynamical 
limit can be applied only for systems in equilibrium, so that an approach based on  statistical 
mechanics  can be reasonably considered. 
However  a dynamical space-time inside a box
is, in general, an out-of-equilibrium system and this fact could lead to misleading results.
In  the next section, we will show a simple argument leading to the conclusion that
black hole evaporation is suppressed by the increasing of the Nariai horizon 
in $f(R)$ gravity. As a consequence, a thermal equilibrium  in an external 
ideal box at $T_{BH}$ will never be reached by a dynamical Nariai black hole. 

\subsection{No particles emission}

Let us consider a Bekenstein-Hawking pair in a dynamical horizon. These are created nearby black hole horizon
and they become real in the external gravitational background. 
Now, one of this pair can pass the horizon as a quantum tunnel effect,
with a certain rate $\Gamma_{bh}$. However, the horizon is displacing outward 
the previous radius because of antievaporation effect. 
As a consequence, the Bekenstein-Hawking pair will be trapped in the black hole interior, 
 in a space-like surface $\mathcal{A}_{space-like}$. From, such a space-like surface, 
the tunneling effect of a particle is impossible. 
As a consequence, the only way to escape is 
if $\Gamma_{bh}^{-1}<\Delta t$, where 
$\Delta t$ is the minimal effective time scale
(from an external observer in a rest frame)
from a $\mathcal{A}_{time-like}\rightarrow \mathcal{A}_{space-like}$
transition -
from a surface on the black hole horizon $\mathcal{A}_{time-like}$ to a surface inside the black hole horizon $\mathcal{A}_{space-like}$.
However, $\Delta t$ can also be infinitesimal of the order of $\lambda$,
where $\lambda$ is the effective separation scale between 
the Bekenstein-Hawking pair. In fact, defining $\Delta r$ as the radius increasing 
with $\Delta t$, it is sufficient $\Delta r>\lambda$ in order to 
"eat" the Bekenstein-Hawking pair in the space-like interior. 
But, for black holes with a radius $r_{S}>>l_{Pl}$,
the tunneling time is expected to be $\Gamma_{bh}^{-1}>>>\Delta t$. 
As a consequence, a realistic Bekenstein-Hawking emission is impossible for 
non-Planckian black holes. 
The same argument can be iteratively applied during all the 
evolution time and the external horizon. 
 In \cite{Ellis:2013oka,Firouzjaee:2014zfa,Firouzjaee:2015bqa}, it was rigorously proven that 
the Bekenstein-Hawking radiation cannot be emitted from a space-like surface by tunneling approach \cite{tunneling}, eikonal approach 
\cite{eikonal}
and Hawking's original derivation by 
Bogolubov coefficients \cite{Hawking}. 

Let us consider the situation from the energy conservation 
point of view. 
In stationary black holes, as in Schwarzschild ones in General Relativity,
the black hole horizon is necessary a Killing bifurcation surface.
In fact, one can define two Killing vector fields 
for the interior and the exterior of the black hole. 
In the exterior region, the Killing vector 
$\zeta^{\mu}$ is time-like, 
while in the interior is space-like. 
This aspect is crucially connected to particles energies: 
the energy of a particle is $E=-p_{\mu}\zeta^{\mu}$,
where $p^{\mu}$ is the 4-momentum. 
As a consequence, energy is always $E>0$ outside the horizon. 
while $E<0$ inside the horizon. In the Killing horizon, 
a real particle creation is energetically possible.
On the other hand, in the dynamical case, 
{\it to define a conserved energy of a particle $E$ for a dynamical space-time 
is impossible}, i.e {\it it is impossible to define Killing vector fields in a dynamical space-time}. 
As discussed above, the Bekenstein-Hawking particle-antiparticle pair 
will be displaced inside the horizon in a space-like region. 
The creation of a real particle from a space-like
region is a violation of causality. 
In fact, it is an a-causal exchange of energy,
{\it i.e} of classical information. 
In fact, a particle inside the horizon is inside a
light-cone with a space-like axis.   

As shown in \cite{Ellis:2013oka}, 
one can distinguish marginally
outer trapped 3-surface
emitting Hawking's pair (timelike surface),
from the outer non-emitting one (space-like).
Let us consider the null or optical Raychaudhuri equation 
for null geodesic congruences: 
\be \label{theta}
\dot{\hat{\theta}}=-\hat{\theta}^{2}-2\hat{\sigma}_{ab}\sigma^{ab}+\hat{\omega}_{cd}\hat{\omega}^{cd}-R_{\mu\nu}k^{\mu}k^{\nu}\,,
\ee
where  hat indicates that expansion, shear, twist and vorticity are defined for the transverse directions. Latin indexes are for spatial components.
The Ricci tensor encodes the dynamical proprieties of gravitational  field. For $f(R)$ gravity, it can be easily derived by an algebraic manipulation of  field equations \eqref{EoM}.
Let us also specify that $\dot{\hat{\theta}}=\frac{\partial}{\partial \lambda}\hat{\theta}$, where $\lambda$ is the affine parameter,
while $k^{a}$ is $k^{a}=\frac{dx^{a}}{d\lambda}$, with $k^{2}=0$, and $\hat{\theta}=k^{a}_{;a}$
also defined as the relative variation of the cross sectional, is
$$
\hat{\theta}=2\frac{1}{A}\frac{dA}{d\lambda}\,.$$
From the above definition, one can define an emitting marginally outer 2-surface $\mathcal{A}_{time-like}$
and the non-emitting inner 2-surface $\mathcal{A}_{space-like}$. Let us call the divergence of the outgoing null geodesics 
$\hat{\theta}_{+}$ in a $S^{2}$-surface.
 With the increasing of the  black hole gravitational field, 
$\hat{\theta}_{+}$ is decreasing (light is more bended). 
On the other hand, the divergence
of ingoing null geodesics is $\hat{\theta}_{-}<0$ 
everywhere, while $\hat{\theta}_{+}>0$ for $r>2M$ in the Schwarzschild case. 
The marginally outer trapped 2-surface $\mathcal{A}^{2d}_{space-like}$
is rigorously defined as a space-like 2-sphere with
\be \label{theta}
\hat{\theta}_{+}(\mathcal{A}_{space-like}^{2d})=0
\ee
As mentioned above, in a Schwarzschild black hole,
the radius of the $S^{2}$-sphere $\mathcal{A}_{space-like}^{2d}$
is exactly equal to the Schwarzschild radius. 
As a consequence, $S^{2}$-spheres with smaller radii 
than $r_{S}=2M$ will be trapped surfaces (TS) with 
$\theta(\mathcal{A}^{2d}_{TS})<0$.

From the 2d surfaces, one can construct a generalized 
definition for 3d surfaces. The dynamical horizon 
is a marginally outer trapped 3d surface. 
It is foliated by marginally trapped 2d surfaces. 
In particular, a dynamical horizon can be 
foliated by a chosen family of $S^{2}$ 
with $\theta_{(n)}$ of a null normal vector $m_a$
vanishing while $\theta_{n\neq m}<0$,
 for each $S^{2}$. 
In particular, one can distinguish among 
an emitting marginally outer trapped 3d surface 
$\mathcal{A}_{time-like}^{3d}$
and a non-emitting one 
$\mathcal{A}_{time-like}^{3d}$
by their derivative of $\hat{\theta}_{m}$
 with respect to 
 an ingoing null tangent vector $n_{a}$. 

%2D surface for 3D

\be \label{fds}
\hat{\theta}_{m}(\mathcal{A}^{3d}_{time-like})=0,\,\,\,\,\,\,\partial \hat{\theta}_{m}(\mathcal{A}^{3d}_{time-like})/\partial n^{a}>0
\ee
and  the non-emitting one is define as 
\be \label{sds}
\hat\theta_{m}(\mathcal{A}^{3d}_{space-like})=0,\,\,\,\,\,\,\partial \hat{\theta}_{m}(\mathcal{A}^{3d}_{space-like})/\partial n^{a}<0
\ee
Now, adopting these definitions,  let us demonstrate that the antievaporation will displace 
the emitting marginally trapped 3d surface to 
a non-emitting space-like 3d surface. 
We can consider the Raychaudhuri equation associated to our problem.
Let us suppose an initial condition $\theta(\bar{\lambda})>0$
with $\bar{\lambda}$ an initial value of the affine parameter
$\lambda$. In the antievaporation phenomena, 
the null Raychauduri equation is bounded 
as 
\be \label{boundR}
\frac{d\hat{\theta}}{d\lambda}<-R_{ab}k^{a}k^{b}
\ee
Let us consider such an equation for an infinitesimal 
$\Delta t$. We can expand the Schwarzschild radius in the Nariai space-time according to the definition \eqref{Horizon}  for $f(R)$ gravity in the above O'Hanlon picture. It is
\be r_{S}=\frac{1}{\mathcal{M}}e^{-\phi_{0}}-\frac{1}{\mathcal{M}}\beta^{2}e^{-\phi_{0}}\phi_{0}t^{2}+\frac{1}{6\mathcal{M}}\beta^{4}e^{- \phi_{0}}\phi_{0}(-2+3\phi_{0})t^{4}+O(t^{5})\,.\ee
where we are  considering only the first 0th leading term. 
For  $\lambda>\bar{\lambda}$, it is
$R_{ab}k^{a}k^{b}>C>0$, where $C$ is a constant associated to the 0-th leading order of $R_{ab}k^{a}k^{b}$ with time. 
As a consequence, $\hat{\theta}$ is bounded as
\be \label{rett}
\hat{\theta}(\lambda)<\hat{\theta}(\lambda)+C(\lambda-\bar{\lambda})
\ee
leading to 
$\hat{\theta}(\lambda)<0$ for $\lambda>\lambda_{1}+\hat{\theta}_{1}/C$, 
where $\lambda_{1},\hat{\theta}_{1}$ are defined at a  characteristic time $t_{1}$. 
Even for a small $\Delta t$,
a constant 0th contribution coming from antievaporation
will cause an extra effective focusing term in the Raychauduri equation. 
On the other hand, the dependence of the extra focusing term 
on time is exponentially growing. 
This formalizes the argument given above. 
As a consequence, an emitting marginally trapped 3d surface will 
exponentially evolve to a non-emitting marginally one.
Bekenstein-Hawking emission is completely suppressed by this dynamical evolution
because of space-like surface cannot emit thermal radiation.
It is important to stress that solutions of Raychaudhuri equations are strictly related to the energy conditions.
In $f(R)$ gravity, energy conditions,  like  the null energy condition, are generically not satisfied as shown in \cite{Albareti:2012va,Mimoso:2014ofa}.

\section{Conclusions and outlooks}

In this paper,  Schwarzschild-de Sitter solutions of  $f(R)$-gravity 
 have been  considered in view of studying their thermodynamical properties. 
If the Schwarzschild radius is comparable to the Hubble radius, we are dealing with  Nariai black holes.
In $f(R)$-gravity, Nariai solutions have been  obtained by Nojiri
and Odintsov \cite{Nojiri:2013su}. They have shown that Nariai black holes 
are unstable: they antievaporate since
 their radii exponentially increase with time (measured by an external observer
in an rest frame). 
This phenomenon is standard for a large class of 
$f(R)$ gravity models.
We have shown that 
Bekenstein-Hawking radiation is 
turned off by antievaporation in $f(R)$ gravity.
In case, the dynamical evolution of the space-time 
traps the emitting  surfaces in the black hole
space-like interior before the effective Bekenstein-Hawking  emission time.
In the limit of very slow black hole antievaporation process 
a part of Bekenstein-Hawking radiation can be emitted.
However, the instability is exponentially growing, so that the
Bekenstein-Hawking radiation could be emitted 
with a very slow antievaporation and only during 
a limited black hole story.

This result opens intriguing possibilities  in cosmology.
In fact, a Nariai black hole is a realistic solution 
for primordial black holes, where the Hubble 
radius is comparable to the primordial black hole radiius
In this case, a black hole  becomes a  Nariai-like solution. 
$f(R)$-gravity has its most successful applications
in cosmology, as a natural and elegant extension of General Relativity, 
providing a successful inflation mechanism 
and reconstructing an effective cosmological term; 
without the introduction 
 of any extra inflaton field
\cite{Nojiri:2006ri,Nojiri:2010wj,Nojiri:2003ft,Capozziello:2003tk,Capozziello:2002rd,Bamba:2012cp,Capozziello:2011et,Capozziello:2007ec,Capozziello:2006dj,Capozziello:2005ku,Capozziello:2009nq,Carloni:2004kp,Carloni:2004kp,Nojiri:2013ru}.
$f(R)$-gravity seems also in good agreement with Planck data and large scale structure,
\cite{DeMartino:2014haa,deMartino:2015zsa,Geng:2015fla}.
 As a consequence, 
to study its primordial or final black hole solutions is crucially important. 
In particular, Nariai primordial black holes present  
 antievaporation instability which allow them to
decay into one (or more than one) final stable vacuum state(s) that can be 
de-Sitter vacua. 
Bekenstein-Hawking radiation cannot reduce the black hole  mass, but gravitational instantons 
can destabilize Nariai black holes, opening wormholes among the Euclidean Schwarzschild-de-Sitter space-time.

In principle, the  transition probability can be evaluated.
A gravitational instanton associated to such a transition has a geometrical action 
$I_{S}=\pi M_{Pl}^{2}R^{2}$, 
where $R=2MM_{Pl}^{-2}=2G_{N}M$
is the radius, 
$M$ black hole mass, 
so that $I_{S}=4\pi M^{2}/M_{Pl}^{2}$. 
Using a diluite instanton gas approximation, one 
can estimate a transition rate for  volume unit 
as $\Gamma=Ae^{-I_{S}}$, where $A\simeq O(1) M^{-1}M_{Pl}^{5}$.
So, the decay rate has a form 
$\Gamma \simeq O(1)M_{Pl}^{5}M^{-1}exp(-4\pi M^{2}/M_{Pl}^{2})$.
For a large black hole $M\simeq M_{\odot}$, 
$\Gamma^{-1}$ is much higher than the whole  universe time-life.
However, for primordial black holes of  mass $M\simeq M_{Pl}$, 
one can estimate a decay rate that is only 
$\Gamma\sim 10^{-6}$. 
One can also consider multi-wormholes' transitions 
from an Euclidean Nariai solution to multi white holes/de Sitter vacua.
In fact, one can "glue" the Penrose diagram of a Nariai solution 
of radius $R$ with Penrose diagrams of $N$ white holes/de-Sitter of radius $R/N$.
This fact could have intriguing connections with cyclic ekpyrotic cosmology:
a Big Crunch/Big Bang transition can be viewed as
a Nariai black hole decay into a white hole/de Sitter vacuum 
However, this is a violation of the null energy condition at Planck scale, 
that, in general,  is dynamically violated in  $f(R)$ gravity. See for example \cite{Battarra:2014naa} for a  realization of a wormhole 
solution in $f(R)$-gravity, violating null energy conditions.
$f(R)$-gravity can naturally be related to 
 cyclic ekpyrotic cosmologies
 \cite{Bamba:2013fha,Odintsov:2015uca,Odintsov:2015uca,Oikonomou:2014jua,Oikonomou:2014yua,Bamba:2014zoa,Odintsov:2014gea,Odintsov:2015zza,Nojiri:2011kd}, where a new universe 
 emerges and inflates from the final Big Crunch
\cite{classici}. Such cyclic Big Bounce is strongly motivated by quantum gravity 
approaches as string theory and loop quantum gravity
See for example
  \cite{Biswas:2005qr,Gasperini:2004ss,Gasperini:2007vw,McAllister:2007bg}.
 \footnote{Reference \cite{Biswas:2005qr} is also related on non-local quantum field theory issues.
 Recently a-causal divergences in non-local supersimmetric scattering amplitudes have been classified in  \cite{Addazi:2015dxa, Addazi:2015ppa}.}.
 Let us also note that in string theory, the presence of higher derivative 
 terms, extending the Einstein-Hilbert action, can be generated by stringy instantons.
 For a review on stringy instantons see
 \cite{Bianchi:2009ij}. Stringy instantons have also other intriguing implications in particle physics as discussed in 
\cite{Addazi:2014ila,Addazi:2014mga,Addazi:2015ata,Addazi:2015rwa,Addazi:2015dxa,
  Addazi:2015hka,Addazi:2015eca,Addazi:2015fua,Addazi:2015oba,Addazi:2015pia,Addazi:2015yna,Addazi:2015ewa}.
Finally, if a classical Nariai black hole is reinterpreted as an ensemble of horizonless 
naked singularities, as recently suggested in   \cite{Addazi:2015gna,Addazi:2015hpa,Addazi:2015bee}, 
one could relax conditions considered in the Bousso-Hawking-Nojiri-Odintsov  analysis.
This hypothesis deserves future investigations both from thermodynamical and cosmological points of view.

\begin{acknowledgments} 
AA would like to thank Gia Dvali for several discussions on quantum black holes 
and his hospitality in Ludwig-Maximilians-Universit$\ddot{a}$t (M$\ddot{u}$nchen) during the preparation of this paper. 
The  work by AA is partially supported  by the MIUR research
grant "Theoretical Astroparticle Physics" PRIN 2012CPPYP7 and by SdC Progetto speciale Multiasse "La Societ\'a della Conoscenza" in Abruzzo PO FSE Abruzzo 2007-2013. SC acknowledges the INFN {\it iniziative specifiche} TEONGRAV and QGSKY.

\end{acknowledgments}


\vspace{0.5cm}

\begin{thebibliography}{99}


%\cite{Nojiri:2013su}
\bibitem{Nojiri:2013su}
  S.~Nojiri and S.~D.~Odintsov,
  %``Anti-Evaporation of Schwarzschild-de Sitter Black Holes in $F(R)$ gravity,''
  Class.\ Quant.\ Grav.\  {\bf 30} (2013) 125003
  [arXiv:1301.2775 [hep-th]].
  %%CITATION = ARXIV:1301.2775;%%
  %14 citations counted in INSPIRE as of 10 Aug 2015

%\cite{Nojiri:2014jqa}
\bibitem{Nojiri:2014jqa}
  S.~Nojiri and S.~D.~Odintsov,
  %``Instabilities and anti-evaporation of ReissnerÐNordstrm black holes in modified $F(R)$ gravity,''
  Phys.\ Lett.\ B {\bf 735} (2014) 376
  [arXiv:1405.2439 [gr-qc]].
  %%CITATION = ARXIV:1405.2439;%%
  %8 citations counted in INSPIRE as of 10 Aug 2015
  
  %\cite{Bousso:1997wi}
\bibitem{Bousso:1997wi}
  R.~Bousso and S.~W.~Hawking,
  %``(Anti)evaporation of Schwarzschild-de Sitter black holes,''
  Phys.\ Rev.\ D {\bf 57} (1998) 2436
  [hep-th/9709224].
  %%CITATION = HEP-TH/9709224;%%
  %96 citations counted in INSPIRE as of 11 Aug 2015
  

%\cite{Katsuragawa:2014hda}
\bibitem{Katsuragawa:2014hda}
  T.~Katsuragawa and S.~Nojiri,
  %``Stability and antievaporation of the SchwarzschildÐde Sitter black holes in bigravity,''
  Phys.\ Rev.\ D {\bf 91} (2015) 8,  084001
  [arXiv:1411.1610 [hep-th]].
  %%CITATION = ARXIV:1411.1610;%%
  %2 citations counted in INSPIRE as of 10 Aug 2015
  
\bibitem{Bekenstein}
J. D. Bekenstein, Phys. Rev. {\bf D7} (1973) 2333.  
  
\bibitem{Hawking}
S. W. Hawking, %{\it Black holes and Thermodynamics}, 
Phys. Rev. D {\bf 13}, 2 (1976).



\bibitem{tunneling}
M. K. Parikh and F. Wilczek, Phys. Rev. Lett. {\bf 85}, 5042 (2000).

\bibitem{eikonal}
C. Barcelo, S. Liberati, S. Sonego and M. Visser, JHEP {\bf 1102}, 003 (2011); \\
C. Barcelo, S. Liberati, S. Sonego and M. Visser, Phys. Rev. D {\bf 83}, 041501 (2011);\\
M. Visser, Int. J. Mod. Phys. D {\bf 12}, 649 (2003)


%\cite{Ellis:2013oka}
\bibitem{Ellis:2013oka}
  G.~F.~R.~Ellis,
  %``Astrophysical black holes may radiate, but they do not evaporate,''
  arXiv:1310.4771 [gr-qc].
  %%CITATION = ARXIV:1310.4771;%%
  %12 citations counted in INSPIRE as of 10 Aug 2015

%\cite{Firouzjaee:2014zfa}
\bibitem{Firouzjaee:2014zfa}
  J.~T.~Firouzjaee and G.~F.~R.~Ellis,
  %``Cosmic Matter Flux May Turn Hawking Radiation Off,''
  Gen.\ Rel.\ Grav.\  {\bf 47} (2015) 2,  6
  [arXiv:1408.0778 [gr-qc]].
  %%CITATION = ARXIV:1408.0778;%%
  %5 citations counted in INSPIRE as of 10 Aug 2015

%\cite{Firouzjaee:2015bqa}
\bibitem{Firouzjaee:2015bqa}
  J.~T.~Firouzjaee and G.~F.~R.~Ellis,
  %``Particle creation from the quantum stress tensor,''
  Phys.\ Rev.\ D {\bf 91} (2015) 10,  103002
  [arXiv:1503.05020 [gr-qc]].
  %%CITATION = ARXIV:1503.05020;%%
  %1 citations counted in INSPIRE as of 10 Aug 2015


  \bibitem{Euclidean}
  G. W. Gibbons, S. W. Hawking, M. J. Perry, Nucl. Phys. B {\bf 138} 141 (1978);
   J. Hartle, K.Schleich, Phys. Rev. D {\bf 36} 2342 (1987).
   
  %\cite{Dyer:2008hb}
\bibitem{Dyer:2008hb}
  E.~Dyer and K.~Hinterbichler,
  %``Boundary Terms, Variational Principles and Higher Derivative Modified Gravity,''
  Phys.\ Rev.\ D {\bf 79} (2009) 024028
  [arXiv:0809.4033 [gr-qc]].
  %%CITATION = ARXIV:0809.4033;%%
  %53 citations counted in INSPIRE as of 11 Aug 2015 
   
   \bibitem{Briscese:2007cd}
  F.~Briscese and E.~Elizalde,
  %``Black hole entropy in modified gravity models,''
  Phys.\ Rev.\ D {\bf 77} (2008) 044009
  [arXiv:0708.0432 [hep-th]].
  %%CITATION = ARXIV:0708.0432;%%
  %46 citations counted in INSPIRE as of 11 Aug 2

          
%ENERGY CONDITIONS

%\cite{Albareti:2012va}
\bibitem{Albareti:2012va}
  F.~D.~Albareti, J.~A.~R.~Cembranos, A.~de la Cruz-Dombriz and A.~Dobado,
  %``On the non-attractive character of gravity in f(R) theories,''
  JCAP {\bf 1307} (2013) 009
  [arXiv:1212.4781 [gr-qc]].
  %%CITATION = ARXIV:1212.4781;%%
  %15 citations counted in INSPIRE as of 11 Aug 2015

%\cite{Mimoso:2014ofa}
\bibitem{Mimoso:2014ofa}
%\cite{Capozziello:2014bqa}
%\item%{Capozziello:2014bqa}
%{\bf ``Generalized energy conditions in Extended Theories of Gravity''}
 S.~Capozziello, F.~S.~N.~Lobo and J.~P.~Mimoso.
 Phys. Rev D {\bf 91}.124019 (2014), arXiv:1407.7293 [gr-qc]\,.
 %\\{}Phys.\ Rev.\ D {\bf 91}, no. 12, 124019 (2015) %(Jul 27, 2014)
%\href{http://inspirehep.net/record/1308477}{HEP entry}
%11 citations counted in INSPIRE as of 19 Jan 2016


%  J.~P.~Mimoso, F.~S.~N.~Lobo and S.~Capozziello,
%  %``Extended Theories of Gravity with Generalized Energy Conditions,''
%  J.\ Phys.\ Conf.\ Ser.\  {\bf 600} (2015) 1,  012047
%  [arXiv:1412.6670 [gr-qc]].
%  %%CITATION = ARXIV:1412.6670;%%

  
%ODINTSOV

%\cite{Nojiri:2006ri}
\bibitem{Nojiri:2006ri}
  S.~Nojiri and S.~D.~Odintsov,
  %``Introduction to modified gravity and gravitational alternative for dark energy,''
  eConf C {\bf 0602061} (2006) 06
   [Int.\ J.\ Geom.\ Meth.\ Mod.\ Phys.\  {\bf 4} (2007) 115]
  [hep-th/0601213].
  %%CITATION = HEP-TH/0601213;%%
  %1386 citations counted in INSPIRE as of 11 Aug 2015  
  
  %\cite{Nojiri:2010wj}
\bibitem{Nojiri:2010wj}
  S.~Nojiri and S.~D.~Odintsov,
  %``Unified cosmic history in modified gravity: from F(R) theory to Lorentz non-invariant models,''
  Phys.\ Rept.\  {\bf 505} (2011) 59
  [arXiv:1011.0544 [gr-qc]].
  %%CITATION = ARXIV:1011.0544;%%
  %945 citations counted in INSPIRE as of 11 Aug 2015
  
  %\cite{Nojiri:2003ft}
\bibitem{Nojiri:2003ft}
  S.~Nojiri and S.~D.~Odintsov,
  %``Modified gravity with negative and positive powers of the curvature: Unification of the inflation and of the cosmic acceleration,''
  Phys.\ Rev.\ D {\bf 68} (2003) 123512
  [hep-th/0307288].
  %%CITATION = HEP-TH/0307288;%%
  %1039 citations counted in INSPIRE as of 11 Aug 2015
  
    
  
%CAPOZZIELLO

%classici
%\cite{Capozziello:2003tk}
\bibitem{Capozziello:2003tk}
  S.~Capozziello, S.~Carloni and A.~Troisi,
  %``Quintessence without scalar fields,''
  Recent Res.\ Dev.\ Astron.\ Astrophys.\  {\bf 1} (2003) 625
  [astro-ph/0303041].
  %%CITATION = ASTRO-PH/0303041;%%
  %566 citations counted in INSPIRE as of 11 Aug 2015
  
  %\cite{Capozziello:2002rd}
\bibitem{Capozziello:2002rd}
  S.~Capozziello,
  %``Curvature quintessence,''
  Int.\ J.\ Mod.\ Phys.\ D {\bf 11} (2002) 483
  [gr-qc/0201033].
  %%CITATION = GR-QC/0201033;%%
  %521 citations counted in INSPIRE as of 11 Aug 2015
  
  %\cite{Bamba:2012cp}
\bibitem{Bamba:2012cp}
  K.~Bamba, S.~Capozziello, S.~Nojiri and S.~D.~Odintsov,
  %``Dark energy cosmology: the equivalent description via different theoretical models and cosmography tests,''
  Astrophys.\ Space Sci.\  {\bf 342} (2012) 155
  [arXiv:1205.3421 [gr-qc]].
  %%CITATION = ARXIV:1205.3421;%%
  %408 citations counted in INSPIRE as of 11 Aug 2015
  
  %\cite{Capozziello:2011et}
\bibitem{Capozziello:2011et}
  S.~Capozziello and M.~De Laurentis,
  %``Extended Theories of Gravity,''
  Phys.\ Rept.\  {\bf 509} (2011) 167
  [arXiv:1108.6266 [gr-qc]].
  %%CITATION = ARXIV:1108.6266;%%
  %459 citations counted in INSPIRE as of 11 Aug 2015
  
  %\cite{Capozziello:2007ec}
\bibitem{Capozziello:2007ec}
  S.~Capozziello and M.~Francaviglia,
  %``Extended Theories of Gravity and their Cosmological and Astrophysical Applications,''
  Gen.\ Rel.\ Grav.\  {\bf 40} (2008) 357
  [arXiv:0706.1146 [astro-ph]].
  %%CITATION = ARXIV:0706.1146;%%
  %443 citations counted in INSPIRE as of 11 Aug 2015
  
    %\cite{Capozziello:2006dj}
\bibitem{Capozziello:2006dj}
  S.~Capozziello, S.~Nojiri, S.~D.~Odintsov and A.~Troisi,
  %``Cosmological viability of f(R)-gravity as an ideal fluid and its compatibility with a matter dominated phase,''
  Phys.\ Lett.\ B {\bf 639} (2006) 135
  [astro-ph/0604431].
  %%CITATION = ASTRO-PH/0604431;%%
  %365 citations counted in INSPIRE as of 11 Aug 2015
  
 %\cite{Capozziello:2005ku}
\bibitem{Capozziello:2005ku} 
  S.~Capozziello, V.~F.~Cardone and A.~Troisi,
  %``Reconciling dark energy models with f(R) theories,''
  Phys.\ Rev.\ D {\bf 71}, 043503 (2005)
  [astro-ph/0501426].
  %%CITATION = ASTRO-PH/0501426;%%
  %280 citations counted in INSPIRE as of 11 Aug 2015
  

\bibitem{Capozziello:2011wg}
  S.~Capozziello and D.~Saez-Gomez,
%  %``Scalar-tensor representation of $f(R)$ gravity and Birkhoff's theorem,''
  Annalen Phys.  {\bf 524} (2012) 279
% % doi:10.1002/andp.201100244
  [arXiv:1107.0948 [gr-qc]].
%  %%CITATION = doi:10.1002/andp.201100244;%%
%  %14 citations counted in INSPIRE as of 17 Jan 2016
%  
  
  
  %\cite{Capozziello:2009nq}
\bibitem{Capozziello:2009nq}
  S.~Capozziello, M.~De Laurentis and V.~Faraoni,
  %``A Bird's eye view of f(R)-gravity,''
  Open Astron.\ J.\  {\bf 3} (2010) 49
  [arXiv:0909.4672 [gr-qc]].
  %%CITATION = ARXIV:0909.4672;%%
  %121 citations counted in INSPIRE as of 11 Aug 2015     
  
%\cite{Carloni:2004kp}
\bibitem{Carloni:2004kp}
  S.~Carloni, P.~K.~S.~Dunsby, S.~Capozziello and A.~Troisi,
  %``Cosmological dynamics of R**n gravity,''
  Class.\ Quant.\ Grav.\  {\bf 22} (2005) 4839
  [gr-qc/0410046].
  %%CITATION = GR-QC/0410046;%%
  %191 citations counted in INSPIRE as of 11 Aug 2015


%f(R) da Planck e cosmologia
%\cite{DeMartino:2014haa}
\bibitem{DeMartino:2014haa}
  I.~De Martino, M.~De Laurentis, F.~Atrio-Barandela and S.~Capozziello,
  %``Probing $f(R)$ gravity with PLANCK data on cluster pressure profiles,''
  J.\ Phys.\ Conf.\ Ser.\  {\bf 600} (2015) 1,  012048
  [arXiv:1412.0095 [astro-ph.CO]].
  %%CITATION = ARXIV:1412.0095;%%

%\cite{deMartino:2015zsa}
\bibitem{deMartino:2015zsa}
  I.~de Martino, M.~De Laurentis and S.~Capozziello,
  %``Constraining $f(R)$ gravity by the Large Scale Structure,''
  Universe 2015, 1(2), 123-157
  [arXiv:1507.06123 [gr-qc]].
  %%CITATION = ARXIV:1507.06123;%%
  




%SARIDAKIS


%\cite{Nojiri:2013ru}
\bibitem{Nojiri:2013ru}
  S.~Nojiri and E.~N.~Saridakis,
  %``Phantom without ghost,''
  Astrophys.\ Space Sci.\  {\bf 347} (2013) 221
  [arXiv:1301.2686 [hep-th]].
  %%CITATION = ARXIV:1301.2686;%%
  %12 citations counted in INSPIRE as of 11 Aug 2015
  
  %\cite{Geng:2015fla}
\bibitem{Geng:2015fla}
  C.~Q.~Geng, M.~W.~Hossain, R.~Myrzakulov, M.~Sami and E.~N.~Saridakis,
  %``Quintessential inflation with canonical and noncanonical scalar fields and Planck 2015 results,''
  Phys.\ Rev.\ D {\bf 92} (2015) 2,  023522
  [arXiv:1502.03597 [gr-qc]].
  %%CITATION = ARXIV:1502.03597;%%
  %3 citations counted in INSPIRE as of 11 Aug 2015
  
  




% CYCLIC UNIVERSE in f(R)

%\cite{Bamba:2013fha}
\bibitem{Bamba:2013fha}
  K.~Bamba, A.~N.~Makarenko, A.~N.~Myagky, S.~Nojiri and S.~D.~Odintsov,
  %``Bounce cosmology from $F(R)$ gravity and $F(R)$ bigravity,''
  JCAP {\bf 1401} (2014) 008
  [arXiv:1309.3748 [hep-th]].
  %%CITATION = ARXIV:1309.3748;%%
  %50 citations counted in INSPIRE as of 11 Aug 2015

%\cite{Odintsov:2015uca}
\bibitem{Odintsov:2015uca}
  S.~D.~Odintsov, V.~K.~Oikonomou and E.~N.~Saridakis,
  %``Superbounce and Loop Quantum Ekpyrotic Cosmologies from Modified Gravity: $F(R)$, $F(G)$ and $F(T)$ Theories,''
  arXiv:1501.06591 [gr-qc].
  %%CITATION = ARXIV:1501.06591;%%
  %6 citations counted in INSPIRE as of 11 Aug 2015
  
  %\cite{Oikonomou:2014jua}
\bibitem{Oikonomou:2014jua}
  V.~K.~Oikonomou,
  %``Loop Quantum Cosmology Matter Bounce Reconstruction from $F(R)$ Gravity Using an Auxiliary Field,''
  arXiv:1412.8195 [gr-qc].
  %%CITATION = ARXIV:1412.8195;%%
  %2 citations counted in INSPIRE as of 11 Aug 2015

%\cite{Oikonomou:2014yua}
\bibitem{Oikonomou:2014yua}
  V.~K.~Oikonomou,
  %``Superbounce and Loop Quantum Cosmology Ekpyrosis from Modified Gravity,''
  arXiv:1412.4343 [gr-qc].
  %%CITATION = ARXIV:1412.4343;%%
  %6 citations counted in INSPIRE as of 11 Aug 201
  
  %\cite{Bamba:2014zoa}
\bibitem{Bamba:2014zoa}
  K.~Bamba, A.~N.~Makarenko, A.~N.~Myagky and S.~D.~Odintsov,
  %``Bounce universe from string-inspired Gauss-Bonnet gravity,''
  JCAP {\bf 1504} (2015) 04,  001
  [arXiv:1411.3852 [hep-th]].
  %%CITATION = ARXIV:1411.3852;%%
  %4 citations counted in INSPIRE as of 11 Aug 2015
  
  %\cite{Odintsov:2014gea}
\bibitem{Odintsov:2014gea}
  S.~D.~Odintsov and V.~K.~Oikonomou,
  %``Matter Bounce Loop Quantum Cosmology from $F(R)$ Gravity,''
  Phys.\ Rev.\ D {\bf 90} (2014) 12,  124083
  [arXiv:1410.8183 [gr-qc]].
  %%CITATION = ARXIV:1410.8183;%%
  %16 citations counted in INSPIRE as of 11 Aug 2015
  
  %\cite{Odintsov:2015zza}
\bibitem{Odintsov:2015zza}
  S.~D.~Odintsov and V.~K.~Oikonomou,
  %``Bouncing cosmology with future singularity from modified gravity,''
  Phys.\ Rev.\ D {\bf 92} (2015) 2,  024016
  [arXiv:1504.06866 [gr-qc]].
  %%CITATION = ARXIV:1504.06866;%%
  %2 citations counted in INSPIRE as of 11 Aug 2015
  
  %\cite{Nojiri:2011kd}
\bibitem{Nojiri:2011kd}
  S.~Nojiri, S.~D.~Odintsov and D.~Saez-Gomez,
  %``Cyclic, ekpyrotic and little rip universe in modified gravity,''
  AIP Conf.\ Proc.\  {\bf 1458} (2011) 207
  [arXiv:1108.0767 [hep-th]].
  %%CITATION = ARXIV:1108.0767;%%
  %56 citations counted in INSPIRE as of 11 Aug 2015
  
  
  
  
  %CLASSICI DEL CICLICO

\bibitem{classici}
 P.~H.~Frampton,
  %``On Cyclic Universes,''
  astro-ph/0612243;L.~Baum and P.~H.~Frampton,
  %``Entropy of contracting universe in cyclic cosmology,''
  Mod.\ Phys.\ Lett.\ A {\bf 23} (2008) 33
  [hep-th/0703162 [HEP-TH]];
J. Khoury, B. A. Ovrut, P. J. Steinhardt and N. Turok, Phys. Rev. D {\bf 64}, 123522 (2001) [arXiv:hep-th/0103239];
J. Khoury, B. A. Ovrut, P. J. Steinhardt and N. Turok, Phys. Rev. D {\bf 66}, 046005 (2002) [arXiv:hep-th/0109050];
P. J. Steinhardt and N. Turok, Science {\bf 312}, 1180 (2006) [arXiv:astro-ph/0605173];
V.G. Gurzadyan, R.Penrose, 
%"On CCC-predicted concentric low-variance circles in the CMB sky", 
Eur.Phys.J. Plus {\bf 128} (2013) 22. 
  
  
  
  
  
  
  
  
  
  %other bounces
  
  %\cite{Biswas:2005qr}
\bibitem{Biswas:2005qr}
  T.~Biswas, A.~Mazumdar and W.~Siegel,
  %``Bouncing universes in string-inspired gravity,''
  JCAP {\bf 0603} (2006) 009
  [hep-th/0508194].
  %%CITATION = HEP-TH/0508194;%%
  %241 citations counted in INSPIRE as of 11 Aug 2015
  
  %\cite{Gasperini:2004ss}
\bibitem{Gasperini:2004ss}
  M.~Gasperini, M.~Giovannini and G.~Veneziano,
  %``Cosmological perturbations across a curvature bounce,''
  Nucl.\ Phys.\ B {\bf 694} (2004) 206
  [hep-th/0401112].
  %%CITATION = HEP-TH/0401112;%%
  %47 citations counted in INSPIRE as of 11 Aug 2015
  
  %\cite{Gasperini:2007vw}
\bibitem{Gasperini:2007vw}
  M.~Gasperini and G.~Veneziano,
  %``String Theory and Pre-big bang Cosmology,''
  hep-th/0703055.
  %%CITATION = HEP-TH/0703055;%%
  %28 citations counted in INSPIRE as of 11 Aug 2015
  
%\cite{McAllister:2007bg}
\bibitem{McAllister:2007bg}
  L.~McAllister and E.~Silverstein,
  %``String Cosmology: A Review,''
  Gen.\ Rel.\ Grav.\  {\bf 40} (2008) 565
  [arXiv:0710.2951 [hep-th]].
  %%CITATION = ARXIV:0710.2951;%%
  %261 citations counted in INSPIRE as of 11 Aug 2015
  
      
      
      
      %\cite{Bianchi:2009ij}
\bibitem{Bianchi:2009ij}
  M.~Bianchi and M.~Samsonyan,
  %``Notes on unoriented D-brane instantons,''
  Int.\ J.\ Mod.\ Phys.\ A {\bf 24} (2009) 5737
  [arXiv:0909.2173 [hep-th]].
  %%CITATION = ARXIV:0909.2173;%%
  %22 citations counted in INSPIRE as of 12 Aug 2015      
      
      
      
      
      %\cite{Addazi:2014ila}
\bibitem{Addazi:2014ila}
  A.~Addazi and M.~Bianchi,
  %``Neutron Majorana mass from exotic instantons,''
  JHEP {\bf 1412} (2014) 089
  [arXiv:1407.2897 [hep-ph]].
  %%CITATION = ARXIV:1407.2897;%%
  %10 citations counted in INSPIRE as of 24 juin 2015

%\cite{Addazi:2014mga}
\bibitem{Addazi:2014mga}
  A.~Addazi and S.~Capozziello,
  %``External stability for Spherically Symmetric Solutions in Lorentz Breaking Massive Gravity,''
  Int.\ J.\ Theor.\ Phys.\  {\bf 54} (2015) 6,  1818
  [arXiv:1407.4840 [gr-qc]].
  %%CITATION = ARXIV:1407.4840;%%
  %7 citations counted in INSPIRE as of 24 Jun 2015

%\cite{Addazi:2015ata}
\bibitem{Addazi:2015ata}
  A.~Addazi,
  %``ÔExotic vector-like pairÕ of color-triplet scalars,''
  JHEP {\bf 1504} (2015) 153
  [arXiv:1501.04660 [hep-ph]].
  %%CITATION = ARXIV:1501.04660;%%
  %9 citations counted in INSPIRE as of 24 juin 2015

%\cite{Addazi:2015rwa}
\bibitem{Addazi:2015rwa}
  A.~Addazi and M.~Bianchi,
  %``Un-oriented Quiver Theories for Majorana Neutrons,''
  JHEP {\bf 1507} (2015) 144
  doi:10.1007/JHEP07(2015)144
  [arXiv:1502.01531 [hep-ph]].
  %%CITATION = doi:10.1007/JHEP07(2015)144;%%
  %14 citations counted in INSPIRE as of 19 Jan 2016

%\cite{Addazi:2015dxa}
\bibitem{Addazi:2015dxa}
  A.~Addazi and G.~Esposito,
  %``Nonlocal quantum field theory without acausality and nonunitarity at quantum level: is SUSY the key?,''
  Int.\ J.\ Mod.\ Phys.\ A {\bf 30} (2015) 1550103
  [arXiv:1502.01471 [hep-th]].
  %%CITATION = ARXIV:1502.01471;%%
  %7 citations counted in INSPIRE as of 24 juin 2015
  
  %\cite{Addazi:2015hka}
\bibitem{Addazi:2015hka}
  A.~Addazi and M.~Bianchi,
  %``Neutron Majorana mass from Exotic Instantons in a Pati-Salam model,''
  JHEP {\bf 1506} (2015) 012
  [arXiv:1502.08041 [hep-ph]].
  %%CITATION = ARXIV:1502.08041;%%
  %8 citations counted in INSPIRE as of 24 juin 2015
  
%\cite{Addazi:2015eca}
\bibitem{Addazi:2015eca}
  A.~Addazi,
  %``More about Neutron Majorana mass from Exotic Instantons: an alternative mechanism in Low-Scale String theory,''
  arXiv:1504.06799 [hep-ph].
  %%CITATION = ARXIV:1504.06799;%%
  %5 citations counted in INSPIRE as of 24 juin 2015

 %\cite{Addazi:2015fua}
\bibitem{Addazi:2015fua}
  A.~Addazi,
  %``Dynamical R-parity violations from exotic instantons,''
  arXiv:1505.00625 [hep-ph], submitted to  EJTP. 
  %%CITATION = ARXIV:1505.00625;%%
  %4 citations counted in INSPIRE as of 24 Jun 2015 
  
%\cite{Addazi:2015oba}
\bibitem{Addazi:2015oba}
  A.~Addazi,
  %``Neutron-antineutron transition as a test-bed for dynamical CPT violations,''
  arXiv:1505.02080 [hep-ph].
  %%CITATION = ARXIV:1505.02080;%%
  %4 citations counted in INSPIRE as of 24 juin 2015

%\cite{Addazi:2015ppa}
\bibitem{Addazi:2015ppa}
  A.~Addazi,
  %``Unitarization and Causalization of Non-local quantum field theories by Classicalization,''
  arXiv:1505.07357 [hep-th], to appear in Int.\ J.\ Mod.\ Phys.\ A. 
  %%CITATION = ARXIV:1505.07357;%%
  %2 citations counted in INSPIRE as of 24 Jun 2015
  

%\cite{Addazi:2015goa}
\bibitem{Addazi:2015goa}
  A.~Addazi,
  %``Direct generation of a Majorana mass for the Neutron from Exotic Instantons,''
  arXiv:1506.06351 [hep-ph].
  %%CITATION = ARXIV:1506.06351;%%
  
  %\cite{Addazi:2015pia}
\bibitem{Addazi:2015pia}
  A.~Addazi,
  %``Neutron-Antineutron oscillation as a test of a New Interaction,''
  Nuovo Cim.\ C {\bf 038} (2015) 01,  21.
  %%CITATION = NUCIA,C038,21;%%
  
  %\cite{Addazi:2015yna}
\bibitem{Addazi:2015yna}
  A.~Addazi, M.~Bianchi and G.~Ricciardi,
  %``Exotic see-saw mechanism for neutrini and leptogenesis in a Pati-Salam model,''
  arXiv:1510.00243 [hep-ph], to appear on JHEP. 
  %%CITATION = ARXIV:1510.00243;%%
  %5 citations counted in INSPIRE as of 19 Jan 2016
  
  %\cite{Addazi:2015ewa}
\bibitem{Addazi:2015ewa}
  A.~Addazi,
  %``Neutron-Antineutron transitions from exotic instantons: how fast they might be and further implications,''
  arXiv:1510.02911 [hep-ph], to appear in the proceeding of 14th Marcell Grossman meeting 2015, Rome, C15-07-12. 
  %%CITATION = ARXIV:1510.02911;%%
  %3 citations counted in INSPIRE as of 19 Jan 2016
  
%\cite{Battarra:2014naa}
\bibitem{Battarra:2014naa}
  L.~Battarra, G.~Lavrelashvili and J.~L.~Lehners,
  %``Creation of wormholes by quantum tunnelling in modified gravity theories,''
  Phys.\ Rev.\ D {\bf 90} (2014) 12,  124015
  [arXiv:1407.6026 [hep-th]].
  %%CITATION = ARXIV:1407.6026;%%
  %1 citations counted in INSPIRE as of 13 Aug 2015
  
  \bibitem{Addazi:2015gna}
  A.~Addazi,
  %``Quantum chaos inside Black Holes,''
  arXiv:1508.04054 [gr-qc], submitted EJTP. 
  %%CITATION = ARXIV:1508.04054;%%
  
  %\cite{Addazi:2015hpa}
\bibitem{Addazi:2015hpa}
  A.~Addazi,
  %``Aspects of quantum chaos inside Black Holes,''
  arXiv:1510.05876 [gr-qc], to appear in the proceeding of Karl Schwarzschild meeting 2015, Frankfurt (C15-07-20.2). 
  %%CITATION = ARXIV:1510.05876;%%
  %2 citations counted in INSPIRE as of 19 Jan 2016
  
  %\cite{Addazi:2015bee}
\bibitem{Addazi:2015bee}
  A.~Addazi, EJTP {\bf 12}, No. IYL15-34 (2015) 89106
  %``Chaotization inside Quantum Black Holes,''
  arXiv:1510.09128 [gr-qc].
  %%CITATION = ARXIV:1510.09128;%%
  
  
  
\end{thebibliography}
\end{document}